\newcommand{\cC}{\ensuremath{\mathcal{C}}}
\newcommand{\cP}{\ensuremath{\mathcal{P}}}
\newcommand{\cT}{\ensuremath{\mathcal{T}}}

\documentclass[12pt,prd,aps,amssymb,amsmath,tightenlines,showpacs]{revtex4}
\usepackage{psfig}
\usepackage{graphics}
\usepackage[pdftex]{graphicx}

\begin{document}


\title{Ghost Busting:\\ $\cP\cT$-Symmetric Interpretation of the Lee Model}

\author{Carl M.\ Bender, Sebastian F.\ Brandt, Jun-Hua Chen, and Qinghai Wang}

\affiliation{Department of Physics, Washington University, St.\ Louis MO 63130,
USA}

\date{\today}

\begin{abstract}
The Lee model was introduced in the 1950s as an elementary quantum field theory
in which mass, wave function, and charge renormalization could be carried out
exactly. In early studies of this model it was found that there is a critical 
value of $g^2$, the square of the renormalized coupling constant, above which
$g_0^2$, the square of the unrenormalized coupling constant, is {\it negative}.
Thus, for $g^2$ larger than this critical value, the Hamiltonian of the Lee
model becomes non-Hermitian. It was also discovered that in this non-Hermitian
regime a new state appears whose norm is negative. This state is called a {\it
ghost} state. It has always been assumed that in this ghost regime the Lee model
is an unacceptable quantum theory because unitarity appears to be violated.
However, in this regime while the Hamiltonian is not Hermitian, it does possess
$\cP\cT$ symmetry. It has recently been discovered that a non-Hermitian
Hamiltonian having $\cP\cT$ symmetry may define a quantum theory that is
unitary. The proof of unitarity requires the construction of a new
time-independent operator called $\cC$. In terms of $\cC$ one can define a new
inner product with respect to which the norms of the states in the Hilbert space
are positive. Furthermore, it has been shown that time evolution in such a
theory is unitary. In this paper the $\cC$ operator for the Lee model in the
ghost regime is constructed in the $V/N\theta$ sector. It is then shown that the
ghost state has a positive norm and that the Lee model is an acceptable unitary
quantum field theory for all values of $g^2$.
\end{abstract}

\pacs{11.30.Er, 11.10.Gh, 11.10.Lm, 03.65.Ge}

\maketitle

\section{Introduction and Brief Review of the Lee Model}
\label{s1}

In 1954 the Lee model was introduced as a quantum field theory in which mass,
wave function, and charge renormalization could be performed exactly and in
closed form \cite{r1,r2,r3,r4}. The Lee model describes a three-particle
interaction of three spinless particles called $V$, $N$, and $\theta$. The $V$
and $N$ particles are fermions and behave roughly like nucleons, and the $\theta$
particle is a boson and behaves roughly like a pion. The basic assumption of the
model is that a $V$ may emit a $\theta$, but when it does so it becomes an $N$.
Also, an $N$ may absorb a $\theta$, but when it does so it becomes a $V$. These
two processes are summarized by
\begin{equation}
V\rightarrow N\,+\,\theta, \qquad N\,+\,\theta\rightarrow V.
\label{e1}
\end{equation}

The Lee model is solvable because it does not respect crossing symmetry; that
is, the crossed processes $V+{\bar\theta}\rightarrow N$ and $N\rightarrow V+{
\bar\theta}$ are forbidden. Eliminating crossing symmetry makes the model
solvable because it introduces two conservation laws. First, {\bf the number of
$N$ quanta plus the number of $V$ quanta is fixed}. Second, {\bf the number of
$N$ quanta minus the number of $\theta$ quanta is fixed.}

As a result of these two highly constraining conservation laws, the Hilbert
space of states decomposes into an infinite number of noninteracting sectors.
The simplest sector is the vacuum sector. Because of the conservation laws there
are no vacuum graphs and the bare vacuum is the physical vacuum. The next two
sectors are the one-$\theta$-particle and the one-$N$-particle sector. These two
sectors are also trivial because the two conservation laws prevent any dynamical
processes from occurring there. As a result, the masses of the $N$ particle and
the $\theta$ particle are not renormalized; that is, the physical masses of
these particles are the same as their bare masses.

The lowest nontrivial sector is the $V/N\theta$ sector. The physical states in
this sector of the Hilbert space are linear combinations of the bare $V$ and the
bare $N\theta$ states and these states consist of the one-physical-$V$-particle
state and the physical $N$-$\theta$-scattering states. To find these states one
can look for the poles and cuts of Green's functions. (The Feynman diagrams are
merely chains of $N\theta$ bubbles connected by single $V$ lines.) The
renormalization in this sector is easy to perform. Following the conventional
renormalization procedure, one finds that the mass of the $V$ particle is
renormalized; that is, the mass of the physical $V$ particle is different from
its bare mass. Most important, in the Lee model one can calculate the
unrenormalized coupling constant as a function of the renormalized coupling
constant in closed form. There are many ways to define the renormalized coupling
constant. For example, in an actual scattering experiment one could define the
square of the renormalized coupling constant $g^2$ as the value of the $N\theta$
scattering amplitude at threshold.

The higher sectors of the Lee model are difficult to study because the equations
are complicated. However, many papers have been written over the years on
various aspects of the Lee model. For example, Weinberg studied the $VN$ sector
because the $V$ and $N$ particles form a bound state whose properties resemble
those of the deuteron \cite{r5}. Amado and Vaughn carried out detailed studies
of scattering amplitudes in the $V\theta$ sector. These examinations required
the solution of difficult integral equations \cite{r6,r7}. Glaser and K\"all\'en
studied the properties of the physical $V$ particle for the case in which the
mass parameters in the Hamiltonian are chosen so that this particle becomes
unstable \cite{r8}. Bender and Nash examined the asymptotic freedom of the Lee
model \cite{r9}.

\begin{figure}[b!]\vspace{3.3in}
\includegraphics{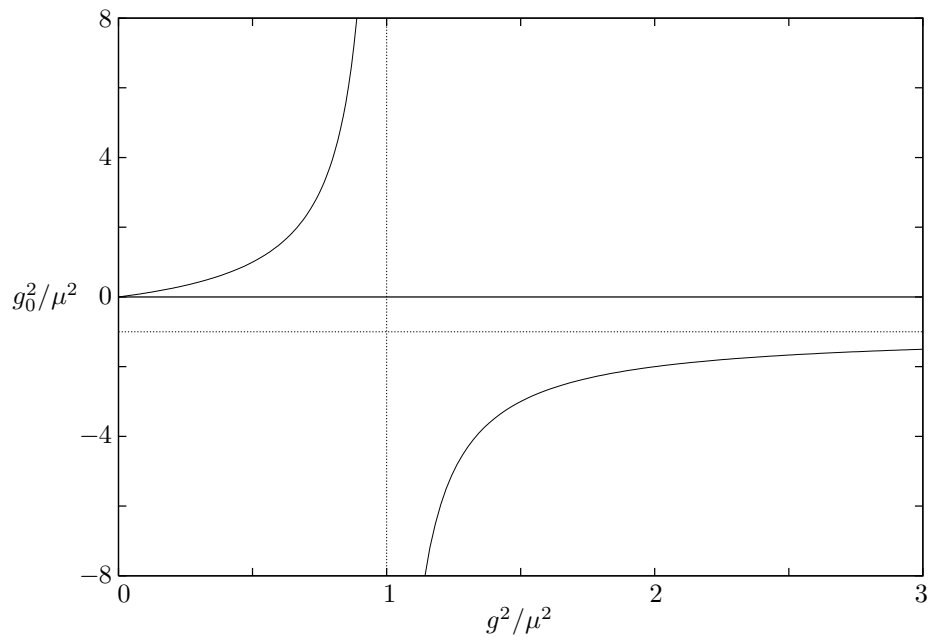}
\caption{Square of the unrenormalized coupling constant, $g_0^2$, plotted as a
function of the square of the renormalized coupling constant $g^2$. When
$g^2=0$ we have $g_0^2=0$, and as $g^2$ increases so does $g_0^2$. However, as
$g^2$ increases past a critical value, $g_0^2$ abruptly becomes negative. In
this regime the unrenormalized coupling constant is negative and the Hamiltonian
is non-Hermitian.}
\label{f1}
\end{figure}

The most interesting aspect of the Lee model is the appearance of a ghost state
in the $V/N\theta$ sector. To understand how this state appears, one must first 
perform coupling-constant renormalization. Expressing $g_0^2$, the square of the
unrenormalized coupling constant, in terms of $g^2$, the square of the
renormalized coupling constant, one obtains a graph like that in Fig.~\ref{f1}.
In principle, the renormalized coupling constant is a physical parameter whose
numerical value is determined by a laboratory experiment. If $g^2$ is measured 
to be near 0, then from Fig.~\ref{f1} we can see that $g_0^2$ is also small.
However, if the laboratory experiment gives a value of $g^2$ that is larger than
this critical value, then the square of the unrenormalized coupling constant
becomes negative. In this regime the unrenormalized coupling constant $g_0$ is 
imaginary and the Hamiltonian is non-Hermitian. Moreover, in this regime a
new state appears in the $V/N\theta$ sector, and because its norm is negative,
the state is called a ghost. There have been numerous attempts to make sense of
the Lee model as a physical quantum theory in the ghost regime \cite{r2,r3,r4}.
However, none of these attempts have been successful. To summarize the
situation, in Ref.~\cite{r4} Barton writes, ``A non-Hermitean Hamiltonian is
unacceptable partly because it may lead to complex energy eigenvalues, but
chiefly because it implies a non-unitary S matrix, which fails to conserve
probability and makes a hash of the physical interpretation.''

The purpose of this paper is to show that, contrary to the view expressed by
Barton, it is indeed possible to give a physical interpretation to what is going
on in the $V/N\theta$ sector of the Lee model when $g_0$ becomes imaginary and
the Hamiltonian becomes non-Hermitian in the Dirac sense. (Dirac Hermitian
conjugation, as indicated by the symbol $\dag$, means combined transpose and
complex conjugate.)

The Lee model is a cubic interaction and there have been several studies of
theories having a cubic interaction multiplied by an imaginary coupling constant
and in all these studies it was found that the spectrum is real and positive.
In two especially important cases it was noticed that the spectrum of Reggeon
field theory is real and positive \cite{r10} and that the spectrum of Lee-Yang
edge-singularity models is also positive \cite{r11}.

The first explanation of how a complex Hamiltonian could have a positive real
spectrum was given by Bender and Boettcher \cite{r12}, who studied the family of
quantum-mechanical Hamiltonians
\begin{equation}
H=p^2+x^2(ix)^\epsilon\quad(\epsilon\geq0).
\label{e2}
\end{equation}
In this study it was shown that these Hamiltonians all have only real, positive,
and discrete energy eigenvalues and it was argued that the unbroken $\cP\cT$
(space-time reflection) symmetry of these Hamiltonians might account for the
reality of the spectrum.

While it is necessary for the spectrum of a Hamiltonian to be real and positive
in order for a Hamiltonian to define a physically acceptable theory of quantum
mechanics, it is not sufficient. One must also have a Hilbert space of states
and an associated inner product whose norm is positive. Positivity of the norm
is essential in order to have a probabilistic interpretation. Moreover, time
evolution must be unitary (probability-conserving). In early attempts to define
the Hilbert space for non-Hermitian $\cP\cT$-symmetric Hamiltonians it was 
conjectured that since the Hermiticity of the Hamiltonian ($H=H^\dag$) was
replaced by $\cP\cT$ symmetry ($H=H^{\cP\cT}$), the Hermitian inner product
$$\langle A|B\rangle\equiv|A\rangle^\dag\cdot|B\rangle$$
should be replaced by the $\cP\cT$ inner product
$$\langle A|B\rangle\equiv |A\rangle^{\cP\cT}\cdot|B\rangle.$$
However, it was quickly observed that while the $\cP\cT$ norm of a state is 
real, the sign of the norm can be either positive or negative, depending on the
state.

A negative norm is physically unacceptable, so it is necessary to find an
alternative definition of the inner product. This definition is given in the
work of Bender, Brody, and Jones \cite{r13}. The key discovery in this paper was
that a $\cP\cT$-symmetric Hamiltonian that has an unbroken $\cP\cT$ symmetry
possesses a new symmetry represented by a linear operator that was called $\cC$.
The $\cC$ operator has three crucial properties. First, the square of the $\cC$
operator is unity,
\begin{equation}
\cC^2=1,
\label{e3a}
\end{equation}
and therefore its eigenvalues are $\pm1$. Second, $\cC$ commutes with $\cP\cT$,
\begin{equation}
\left[\cC,\cP\cT\right]=0,
\label{e3b}
\end{equation}
and therefore $\cC$ is $\cP\cT$ symmetric. Third, $\cC$ commutes with the
Hamiltonian,
\begin{equation}
\left[\cC,H\right]=0,
\label{e3c}
\end{equation}
and therefore the eigenstates of the Hamiltonian are also eigenstates of $\cC$.
In fact, states of $H$ having a negative $\cP\cT$ norm have eigenvalue $-1$
under $\cC$, and eigenstates of $H$ having a positive $\cP\cT$ norm have
eigenvalue $+1$ under $\cC$. From these three properties one can use the $\cC$
operator to construct the correct inner product for the Hilbert space of states
of the Hamiltonian:
\begin{equation}
\langle A|B\rangle\equiv|A\rangle^{\cC\cP\cT}\cdot|B\rangle.
\label{e4}
\end{equation}
This $\cC\cP\cT$ inner product is associated with a {\it positive} norm.
Furthermore, the usual time translation operator $e^{iHt}$ preserves the inner
product. Thus, with respect to this inner product, time evolution is unitary.

The implication of Ref.~\cite{r13} is that a non-Hermitian $\cP\cT$-symmetric
Hamiltonian determines its own Hilbert space of state vectors and the associated
inner product. One can thus regard a non-Hermitian $\cP\cT$-symmetric
Hamiltonian as a {\it bootstrap} theory because one must find the eigenvectors
and eigenvalues of the Hamiltonian in order to discover what the Hilbert space
is.

The key step in understanding non-Hermitian $\cP\cT$-symmetric quantum theories
is constructing the $\cC$ operator. Several papers have been published regarding
the construction of this operator. A perturbative calculation of $\cC$ for cubic
quantum-mechanical theories was given in Ref.~\cite{r14} and a nonperturbative
WKB calculation of $\cC$ for the more general quantum-mechanical theory in
(\ref{e2}) was given in Ref.~\cite{r15}. A perturbative calculation of the $\cC$
operator in $D$-dimensional cubic quantum field theories was done
\cite{r16,r17}. From these papers it is evident that the best way to calculate
$\cC$ in a theory with a cubic interaction is to solve the system of operator
equations in (\ref{e3a}) - (\ref{e3c}).

This paper is organized very simply. In Sec.~\ref{s2} we consider first the
case of a {\it quantum-mechanical} Lee model. It is shown that when the
renormalized coupling constant is larger than the critical value shown in
Fig.~\ref{f1}, the Hamiltonian becomes non-Hermitian and a ghost state appears
in the $V/N\theta$ sector of the theory. In this regime the theory is $\cP\cT$
symmetric. Kleefeld was the first to point out this transition to $\cP\cT$
symmetry \cite{r18}; this reference gives a beautiful and comprehensive review
of the history of non-Hermitian Hamiltonians. The $\cP\cT$ norm of the physical
$V$ particle is positive, but the $\cP\cT$ norm of the ghost is negative. We
must then calculate the $\cC$ operator for this theory, and we do this {\it
exactly and in closed form.} We verify that the $\cC\cP\cT$ norms of the $V$ and
the ghost states are positive. Because the $\cC\cP\cT$ norm of the ghost state
is positive the term ``ghost'' is actually inappropriate. With this metric the
Lee model in this sector is a fully consistent and unitary quantum theory.

Next, in Sec.~\ref{s3} we generalize the results in Sec.~\ref{s2} to the case of
a {\it quantum-field-theoretic} Lee model. Again we verify that when the
renormalized coupling constant is larger than a critical value, the Hamiltonian
becomes non-Hermitian and a ghost state appears in the $V/N\theta$ sector. The
$\cP\cT$ norms of the physical $V$ and the $N$-$\theta$ scattering states are
positive but the $\cP\cT$ norm of the ghost state is negative. As in
Sec.~\ref{s2}, we again calculate the $\cC$ operator in the $V/N\theta$ sector
and we verify that the $\cC\cP\cT$ norms of all states in the $V/N\theta$ sector
are positive. These calculations demonstrate that there is no ghost and that the
Lee model in the $V/N\theta$ sector is a fully consistent unitary quantum
theory.

\section{Quantum-Mechanical Lee Model}
\label{s2}

The quantum-mechanical Lee model describes the interaction of the three
particles $V$, $N$, and $\theta$, but in this model there is only a time
variable and no space variable. Correspondingly, there is only energy and no
momentum. Particles do not move but simply sit at one point and evolve in time.
To create states of these particles we apply the creation operators $V^\dag$,
$N^\dag$, and $a^\dag$ to the vacuum state $|0\rangle$. The bare states in
this model are given by
\begin{eqnarray}
|1,0,0\rangle &\equiv& V^\dag|0\rangle,\nonumber\\
|0,1,0\rangle &\equiv& N^\dag|0\rangle,\nonumber\\
|0,0,1\rangle &\equiv& a^\dag|0\rangle,\nonumber\\
|1,0,n\rangle &\equiv& V^\dag\frac{\left(a^\dag\right)^n}{\sqrt{n!}} 
|0\rangle\quad(n\in{\mathbb Z},~n\geq0),\nonumber\\
|0,1,n\rangle &\equiv& N^\dag\frac{\left(a^\dag\right)^n}{\sqrt{n!}} 
|0\rangle\quad(n\in{\mathbb Z},~n\geq0),\nonumber\\
|1,1,n\rangle &\equiv& V^\dag N^\dag\frac{\left(a^\dag\right)^n}{\sqrt{n!}}
|0\rangle\quad(n\in{\mathbb Z},~n\geq0).
\label{e5}
\end{eqnarray}
We treat $\theta$ as a boson, but treat $V$ and $N$ as fermions, so there are no
multi-$V$ or multi-$N$ states. These creation and annihilation operators satisfy
the following commutation and anticommutation relations:
\begin{eqnarray}
\left[a,a\right]&=&[a^\dag,a^\dag]=0,\quad[a,a^\dag]=1,\nonumber\\
\left[a,N\right]&=&[a,V]=[a,N^\dag]=[a,V^\dag]=[a^\dag,N]=[a^\dag,V]
=[a^\dag,N^\dag]=[a^\dag,V^\dag]=0,\nonumber\\
\left[N,N\right]_+&=&[N^\dag,N^\dag]_+=[V,V]_+=[V^\dag,V^\dag]_+=0,\nonumber\\
\left[V,N\right]_+&=&[V^\dag,N^\dag]_+=[N,V^\dag]_+=[N^\dag,V]_+=0,\nonumber\\
\left[N,N^\dag\right]_+&=&[V,V^\dag]_+=1.
\label{e6}
\end{eqnarray}

The Hamiltonian for the quantum-mechanical Lee model has the form
\begin{equation}
H=H_0+g_0 H_1,
\label{e7}
\end{equation}
where
\begin{eqnarray}
H_0&=&m_{V_0}V^\dag V+m_N N^\dag N+m_\theta a^\dag a,\nonumber\\
H_1&=&V^\dag Na+a^\dag N^\dag V.
\label{e8}
\end{eqnarray}
The bare states (\ref{e5}) are the eigenstates of $H_0$ and the physical states
are the eigenstates of the full Hamiltonian $H$. Note that the mass parameters
$m_N$ and $m_\theta$ represent the {\it physical} masses of the one-$N$-particle
and one-$\theta$-particle states because these states do not undergo mass
renormalization. However, $m_{V_0}$ is the {\it bare} mass of the $V$ particle.

The $V$, $N$, and $\theta$ particles are all treated as pseudoscalars. To
understand why this is so, recall that in quantum mechanics the position
operator
$$x=\frac{1}{\sqrt{2}}(a+a^\dag)$$
and the momentum operator
$$p=\frac{1}{i\sqrt{2}}(a-a^\dag)$$
both change sign under parity reflection:
\begin{equation}
\cP x\cP=-x,\quad\cP p\cP=-p.
\label{e9}
\end{equation}
Thus, we conclude that
\begin{equation}
\cP V\cP=-V,~\cP N\cP=-N,~\cP a\cP=-a,~
\cP V^\dag\cP=-V^\dag,~\cP N^\dag\cP=-N^\dag,~\cP a^\dag\cP=-a^\dag.
\label{e10}
\end{equation}

Under time reversal $p$ and $i$ change sign but $x$ does not:
\begin{equation}
\cT p\cT=-p,\quad\cT i\cT=-i,\quad \cT x\cT=x.
\label{e11}
\end{equation}
Thus,
\begin{equation}
\cT V\cT=V,~\cT N\cT=N,~\cT a\cT=a,~
\cT V^\dag\cT=V^\dag,~\cT N^\dag\cT=N^\dag,~\cT a^\dag\cT=a^\dag.
\label{e12}
\end{equation}

Note that when the bare coupling constant $g_0$ is real, $H$ in (\ref{e7}) is
Hermitian: $H^\dag=H$. When $g_0$ is imaginary,
\begin{equation}
g_0=i\lambda_0\quad(\lambda_0~{\rm real}),
\label{e13}
\end{equation}
$H$ is not Hermitian, but by virtue of the transformation properties in
(\ref{e10}) and (\ref{e12}), $H$ is $\cP\cT$-symmetric: $H^{\cP\cT}=H$.

Let us first assume that $g_0$ is real so that $H$ is Hermitian and let us
examine the simplest nontrivial sector of the quantum-mechanical Lee model;
namely, the $V/N\theta$ sector. To do so, we look for the eigenstates of the
Hamiltonian $H$ in the form of linear combinations of the bare one-$V$-particle
and the bare one-$N$-one-$\theta$-particle states. We find that there are two
eigenfunctions and two eigenvalues. We interpret the eigenfunction corresponding
to the lower-energy eigenvalue as the physical one-$V$-particle state, and we
interpret the eigenfunction corresponding with the higher-energy eigenvalue as
the physical one-$N$-one-$\theta$-particle state. (In the field-theoretic
version of the Lee model this higher-energy state corresponds to the continuum
of physical $N$-$\theta$ scattering states.) Thus, we make the {\it ansatz}
\begin{eqnarray}
|V\rangle&=&c_{11}|1,0,0\rangle+c_{12}|0,1,1\rangle,\nonumber\\
|N\theta\rangle&=&c_{21}|1,0,0\rangle+c_{22}|0,1,1\rangle,
\label{e14}
\end{eqnarray}
and demand that these states be eigenstates of $H$ with eigenvalues $m_V$ (the
renormalized $V$-particle mass) and $E_{N\theta}$. The eigenvalue problem
reduces to a pair of elementary algebraic equations:
\begin{eqnarray}
c_{11}m_{V_0}+c_{12}g_0&=&c_{11}m_V,\nonumber\\
c_{21}g_0+c_{22}\left(m_N+m_\theta\right)&=&c_{22}E_{N\theta}.
\label{e15}
\end{eqnarray}
The solutions to (\ref{e15}) are
\begin{eqnarray}
m_V&=&\frac{1}{2}\left(m_N+m_\theta+m_{V_0}-\sqrt{\mu_0^2+4g_0^2}\right),
\nonumber\\
E_{N\theta}&=&\frac{1}{2}\left(m_N+m_\theta+m_{V_0}+\sqrt{\mu_0^2+4g_0^2}
\right),
\label{e16}
\end{eqnarray}
where
\begin{equation}
\mu_0\equiv m_N+m_\theta-m_{V_0}.
\label{e17}
\end{equation}
Notice that $m_V$, the mass of the physical $V$ particle, is different from
$m_{V_0}$, the mass of the bare $V$ particle, because the $V$ particle undergoes
mass renormalization.

Next, we perform wave-function renormalization. Following Barton we define the 
wave-function renormalization constant $Z_V$ by the relation \cite{r4}
\begin{equation}
1=\langle 0|\frac{1}{\sqrt{Z_V}}V|V\rangle.
\label{e18}
\end{equation}
This gives
\begin{equation}
Z_V=\frac{2g_0^2}{\sqrt{\mu_0^2+4g_0^2}\left(\sqrt{\mu_0^2+4g_0^2}-\mu_0\right)
}.
\label{e19}
\end{equation}

Finally, we perform coupling-constant renormalization. Again, following Barton
we note that $\sqrt{Z_V}$ is the ratio between the renormalized coupling
constant $g$ and the bare coupling constant $g_0$ \cite{r4}. Thus,
\begin{equation}
\frac{g^2}{g_0^2}=Z_V.
\label{e20}
\end{equation}
After some elementary algebra we find that in terms of the renormalized mass and
coupling constant, the bare coupling constant satisfies:
\begin{equation}
g_0^2=\frac{g^2}{1-\frac{g^2}{\mu^2}},
\label{e21}
\end{equation}
where $\mu$ is defined as
\begin{equation}
\mu\equiv m_N+m_\theta-m_V.
\label{e22}
\end{equation}
We cannot freely choose the value of $g$ because the value of $g$ is in
principle taken from experimental data. Once the value of $g$ has been
determined experimentally, we can use (\ref{e21}) to determine $g_0$. The
relation in (\ref{e21}) is plotted in Fig.~\ref{f1}. Figure \ref{f1} reveals an
extremely surprising property of the Lee model: If $g$ is larger than the
critical value $\mu$, then the square of $g_0$ is negative, and $g_0$ is
imaginary.

\begin{figure}[b!]\vspace{3.4in}
\includegraphics{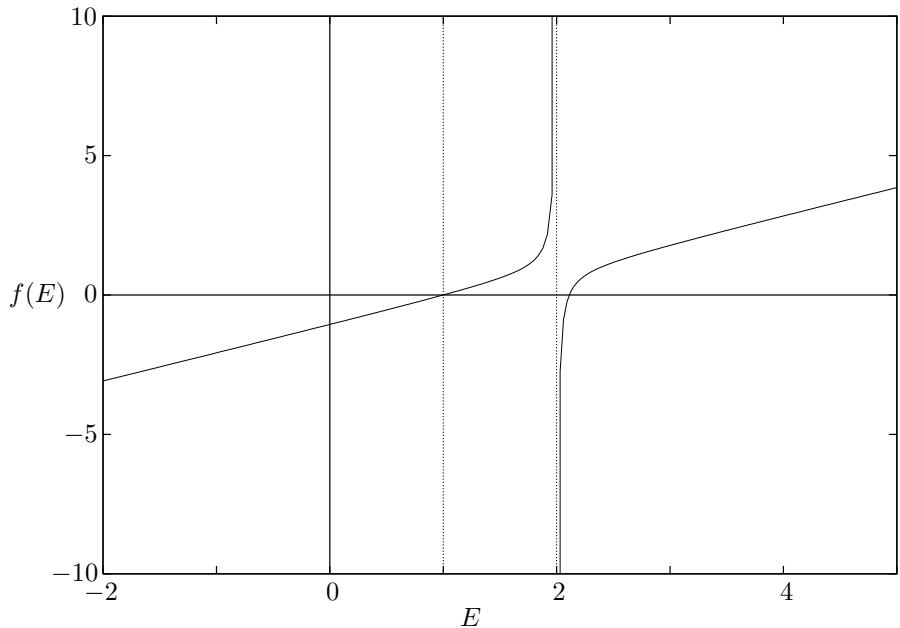}
\caption{Plot of the secular determinant $f(E)$ obtained by applying Cramer's
rule to (\ref{e15}) for $g_0$ real and small. Values of $E$ for which $f(E)=0$
correspond to eigenvalues of the Hamiltonian (\ref{e7}). Observe that $f(E)$ has
two zeros, the lower one corresponding to the energy of the physical $V$
particle and the upper one corresponding to the energy of the physical $N\theta$
state.}
\label{f2}
\end{figure}

\begin{figure}[b!]\vspace{3.4in}
\includegraphics{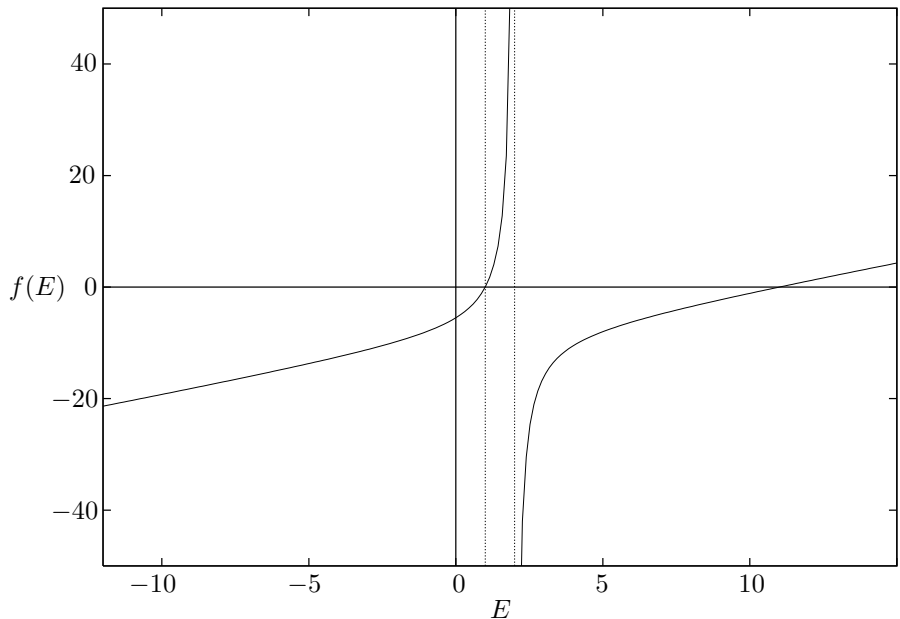}
\caption{Same as in Fig.~\ref{f2} except that the value of $g$ is larger. Note
that the larger eigenvalue $E$ (the larger value of $E$ for which $f(E)=0$),
which corresponds to the physical $N\theta$ state, has moved up the real-$E$
axis.}
\label{f3}
\end{figure}

\begin{figure}[b!]\vspace{3.4in}
\includegraphics{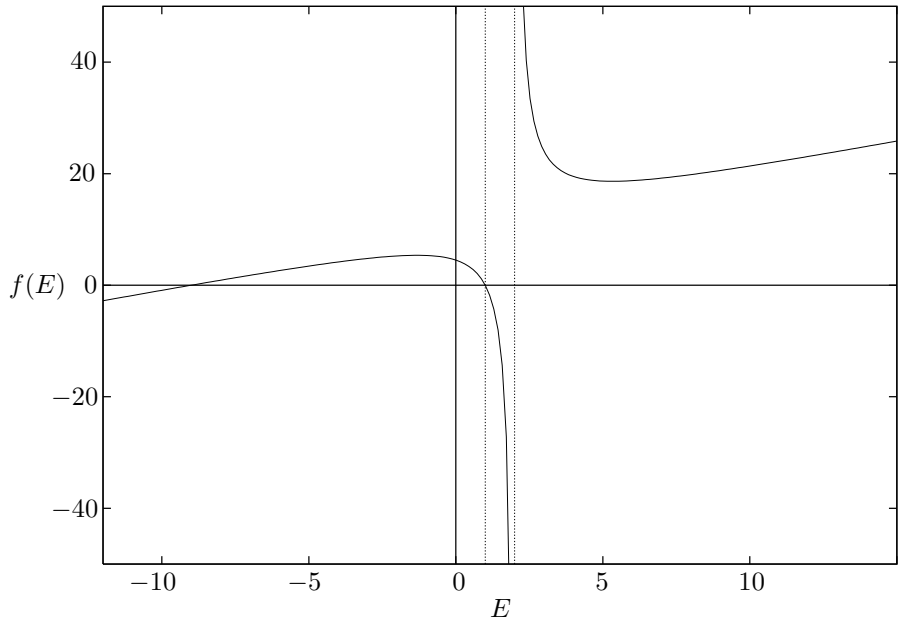}
\caption{Same as in Fig.~\ref{f3} except that $g$ is larger than its critical
value and the unrenormalized coupling constant $g_0$ is imaginary. In this
regime the Hamiltonian is non-Hermitian. Observe that the larger zero of $f(E)$
has moved out to infinity and is now moving up the negative real-$E$ axis below
the energy of the physical $V$ particle. The $N\theta$ state has disappeared and
has been replaced by a ghost state.}
\label{f4}
\end{figure}

\begin{figure}[b!]\vspace{3.4in}
\includegraphics{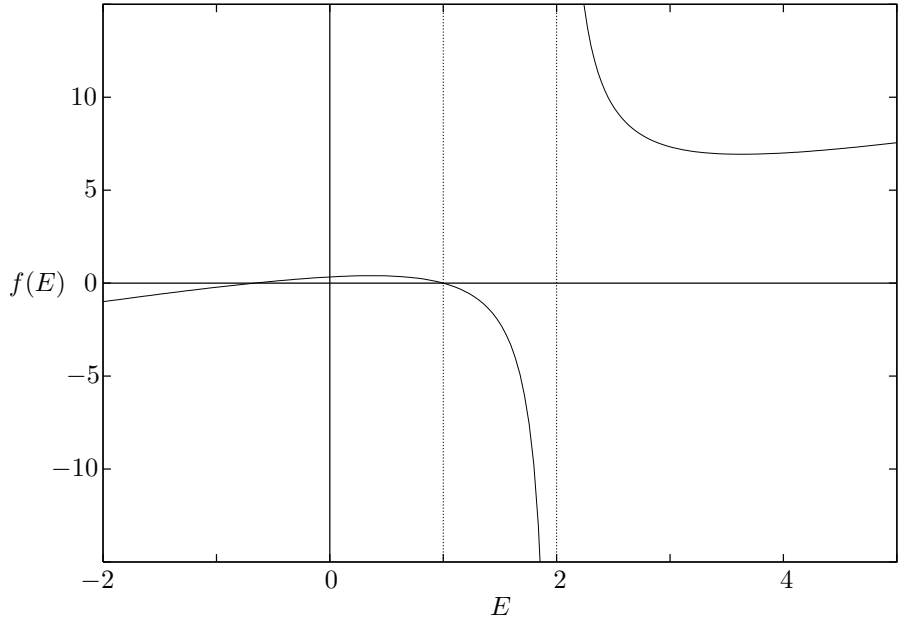}
\caption{Same as in Fig.~\ref{f4} except that $g$ has an even larger value. Note
that as $g$ continues to increase the ghost energy continues to move up towards
the energy of the physical $V$ particle.}
\label{f5}
\end{figure}

As $g$ approaches its critical value from below, the two energy eigenvalues in
(\ref{e16}) vary accordingly. The energy eigenvalues are the two zeros of the
secular determinant $f(E)$ obtained from applying Cramer's rule to (\ref{e15}).
We have plotted $f(E)$ as a function of $E$ in Figs.~\ref{f2} - \ref{f5}. Figure
\ref{f2} shows $f(E)$ for very small $g$ and Fig.~\ref{f3} shows $f(E)$ for a
larger value of $g$, but one for which $g$ is still smaller than its critical
value. As $g$ (and $g_0$) increase, the energy of the physical $N\theta$ state
increases. The energy of the $N\theta$ state becomes infinite as $g$ reaches its
critical value. As $g$ increases past its critical value, the upper energy
eigenvalue goes around the bend: It abruptly jumps from being large and positive
to being large and negative (see Fig.~\ref{f4}). Then, as $g$ continues to
increase, this energy eigenvalue approaches the energy of the physical $V$
particle from below.

When $g$ increases past its critical value, the Hamiltonian $H$ in (\ref{e7})
becomes non-Hermitian, but its eigenvalues in the $V/N\theta$ sector remain
real, as we can see from Figs.~\ref{f4} and \ref{f5}. The eigenvalues remain
real because $H$ becomes $\cP\cT$ symmetric, and cubic $\cP\cT$-symmetric
Hamiltonians that have been studied in the past have been shown to have real
spectra. However, in the $\cP\cT$-symmetric regime it is no longer appropriate
to interpret the lower eigenvalue as the energy of the physical $N\theta$ state.
Rather, it is the energy of a new kind of state $|G\rangle$ called a {\it
ghost}. As is shown in Refs.~\cite{r2,r3,r4}, the Hermitian norm of this state
is {\it negative}. Until the writing of this paper, a satisfactory physical
interpretation of the ghost state had not been found.

We will now show how a physical interpretation of the ghost state emerges easily
when we use the methods developed in Ref.~\cite{r13}. We begin by verifying that
in the $\cP\cT$-symmetric regime, where $g_0$ is imaginary, the states of the
Hamiltonian are eigenstates of the $\cP\cT$ operator, and we then choose the
multiplicative phases of these states so that their $\cP\cT$ eigenvalues are
unity:
$$\cP\cT|G\rangle=|G\rangle,\quad \cP\cT|V\rangle=|V\rangle.$$
It is then straightforward to verify that the $\cP\cT$ norm of the $V$ state
is positive, while the $\cP\cT$ norm of the ghost state is negative.

We then follow the procedures explained in Refs.~\cite{r16,r17} to calculate the
$\cC$ operator. In these papers it is shown that the $\cC$ operator can be
expressed as an exponential of a function $Q$ multiplying the parity operator:
\begin{equation}
\cC=e^{Q(V^\dag,V;N^\dag,N;a^\dag,a)}\cP.
\label{e23}
\end{equation}
We then impose the algebraic operator equations in (\ref{e3a}) - (\ref{e3c}).
The condition $\cC^2=1$ gives
\begin{equation}
Q(V^\dag,V;N^\dag,N;a^\dag,a)=-Q(-V^\dag,-V;-N^\dag,-N;-a^\dag,-a).
\label{e24}
\end{equation}
Thus, $Q(V^\dag,V;N^\dag,N;a^\dag,a)$ is an odd function in total powers
of $V^\dag$, $V$, $N^\dag$, $N$, $a^\dag$, and $a$.

Next, we impose the condition $[\cC,\cP\cT]=0$ and obtain the result that
\begin{equation}
Q(V^\dag,V;N^\dag,N;a^\dag,a)=Q^*(-V^\dag,-V;-N^\dag,-N;-a^\dag,-a),
\label{e25}
\end{equation}
where $*$ denotes complex conjugation.

Finally, we impose the condition that the operator $\cC$ commutes with $H$:
$[\cC, H]=0$, which requires that
\begin{equation}
\left[e^Q, H_0\right]= g_0\left[e^Q,H_1\right]_+.
\label{e26}
\end{equation}
Although in Refs.~\cite{r16,r17} we were only able to find the $\cC$ operator
to leading order in perturbation theory, for the Lee model it is possible to
calculate the $\cC$ operator exactly and in closed form. To do so, we seek a
solution for $Q$ as a formal Taylor series in powers of $g_0$:
\begin{equation}
Q=\sum_{n=0}^{\infty}g_0^{2 n+1}Q_{2n+1}.
\label{e27}
\end{equation}
Only odd powers of $g_0$ appear in this series and $Q_{2n+1}$ are all 
anti-Hermitian: $Q_{2n+1}^\dag=-Q_{2n+1}$. From (\ref{e26}) we get
\begin{eqnarray}
Q_1&=&\frac{2}{\mu_0}\left(V^\dag Na-a^\dag N^\dag V\right),\nonumber\\
Q_3&=&-\frac{8}{3\mu_0^3}\left(V^\dag Nan_\theta-n_\theta a^\dag N^\dag V\right)
,\nonumber\\
Q_5&=&\frac{32}{5\mu_0^5}\left(V^{\dag}Nan_\theta^2-n_\theta^2a^\dag N^\dag V
\right),\nonumber\\
&\vdots&\nonumber\\
Q_{2n+1}&=&(-1)^n\frac{2^{2n+1}}{(2n+1)\mu_0^{2n+1}}\left(V^\dag Nan_\theta^n
-n_\theta^na^\dag N^\dag V\right),
\label{e28}
\end{eqnarray}
and so on, where $n_\theta=a^\dag a$ is the number operator for
$\theta$-particle quanta.

Finally, we sum over all $Q_{2n+1}$ and obtain the {\it exact} result that
\begin{equation}
Q=V^\dag Na\frac{1}{\sqrt{n_\theta}}{\rm arctan}\left(\frac{2g_0\sqrt{n_\theta}}
{\mu_0}\right)-\frac{1}{\sqrt{n_\theta}}{\rm arctan}\left(\frac{2g_0
\sqrt{n_\theta}}{\mu_0}\right) a^\dag N^\dag V.
\label{e29}
\end{equation}
We must now exponentiate this result to obtain the $\cC$ operator. Fortunately,
the exponential of $Q$ simplifies considerably because we are treating the $V$
and $N$ particles as fermions and therefore we can use the identity $n_{V,N}^2=
n_{V,N}$. Our exact result for $e^Q$ is
\begin{eqnarray}
e^Q&=&\left[1-n_V-n_N+n_Vn_N+\frac{\mu_0n_N\left(1-n_V\right)}{\sqrt{\mu_0^2+4
g_0^2n_\theta}}+\frac{\mu_0n_V\left(1-n_N\right)}{\sqrt{\mu_0^2+4g_0^2\left(
n_\theta+1\right)}}\right.\nonumber\\
&&\qquad\left.+V^\dag Na\frac{2g_0\sqrt{n_\theta}}{\sqrt{\mu_0^2+4g_0^2n_\theta}
}-\frac{2g_0\sqrt{n_\theta}}{\sqrt{\mu_0^2+4g_0^2n_\theta}}a^\dag N^\dag V
\right].
\label{e30}
\end{eqnarray}

We can also express the parity operator $\cP$ in terms of number operators:
\begin{equation}
\cP=e^{i\pi\left(n_V+n_N+n_\theta\right)}=\left(1-2n_V\right)\left(1-2n_N
\right)e^{i\pi n_\theta}.
\label{e31}
\end{equation}
Combining $e^Q$ and $\cP$, we obtain the exact expression for $\cC$:
\begin{eqnarray}
\cC&=&\left[1-n_V-n_N+n_Vn_N+\frac{\mu_0n_N\left(1-n_V\right)}{\sqrt{\mu_0^2+4
g_0^2n_\theta}}+\frac{\mu_0n_V\left(1-n_N\right)}{\sqrt{\mu_0^2+4g_0^2\left(
n_\theta+1\right)}}\right.\nonumber\\
&&\qquad\left.+V^\dag Na\frac{2g_0\sqrt{n_\theta}}{\sqrt{\mu_0^2+4g_0^2n_\theta
}}-\frac{2g_0\sqrt{n_\theta}}{\sqrt{\mu_0^2+4g_0^2n_\theta}}a^\dag
N^\dag V\right]\left(1-2n_V\right)\left(1-2n_N\right)e^{i\pi n_\theta}.
\label{e32}
\end{eqnarray}

Using this $\cC$ operator we calculate the $\cC\cP\cT$ norm of the $V$ state and
the ghost state and find that they are both positive. Furthermore, as is shown
in Ref.~\cite{r13}, time evolution is unitary. Thus, we have verified that with
the proper definition of the inner product, the quantum-mechanical Lee model is 
a physically acceptable and fully consistent quantum theory even in the ghost
regime where the unrenormalized coupling constant is imaginary.

\section{Quantum-Field-Theoretic Lee Model}
\label{s3}

The field-theoretic Lee model is more general than the quantum-mechanical Lee
model discussed in Sec.~\ref{s2} in that it allows the $V$, $N$, and $\theta$
particles to move. Thus, the Hamiltonian for the field-theoretic Lee model is
composed of operators that create and annihilate $V$, $N$, and $\theta$
particles of a given momentum. This Hamiltonian has the form
\begin{equation}
H=H_0+H_I.
\label{e33}
\end{equation}
The free Hamiltonian is given by
\begin{equation}
H_0=m_{V_0}\int d\vec{p}\,V^\dag_{\vec p}V_{\vec p}+m_N\int d\vec{p}\,N^\dag_{
\vec p}N_{\vec p}+\int d\vec{k}\,\omega_{\vec k}a^\dag_{\vec k}a_{\vec k},
\label{e34}
\end{equation}
where $\omega_{\vec k}=\sqrt{{\vec k}^2+m_\theta^2}$, and the interaction
Hamiltonian is given by
\begin{equation}
H_I=\int d\vec{k}\,h_{\vec k}\int d\vec{p}\,\left[V^\dag_{\vec p}N_{\vec{p}-
\vec{k}}a_{\vec k}+a^\dag_{\vec k}N^{\dag}_{{\vec p}-{\vec k}}V_{\vec p}\right],
\label{e35}
\end{equation}
where
\begin{equation}
h_{\vec k}=\frac{g_0\rho(\omega_{\vec k})}{(2\pi)^{3/2}\sqrt{2\omega_{\vec k}}}.
\label{e36}
\end{equation}
The function $\rho(\omega_{\vec k})$ is a cut-off that ensures that the process
of renormalization can be carried out using finite quantities. Note that when
the unrenormalized coupling constant $g_0$ is imaginary, $g_0=i\lambda_0$
($\lambda_0$ real), $H$ is not Hermitian but is $\cP\cT$ symmetric.

The $V/N\theta$ sector of the Lee model is spanned by the one-bare-$V$-particle
state and the one-bare-$N$-one-bare-$\theta$-particle states. Because
of the conservation laws of the Lee model, this sector is an invariant subspace.
The physical energy eigenstates in this sector have the general form $|E,\vec{p}
\rangle$, where
\begin{equation}
|E,\vec{p}\rangle=\varepsilon\left(V_{\vec{p}}^\dag+\int d\vec{k}\,\varphi_
{\vec k}N_{\vec{p}-\vec{k}}^\dag a_{\vec{k}}^\dag\right)|0\rangle.
\label{e37}
\end{equation}
Under the action of the Hamiltonian this state has the energy eigenvalue $E$:
\begin{equation}
H|E,\vec{p}\rangle=E|E,\vec{p}\rangle.
\label{e38}
\end{equation}

The eigenvalue problem (\ref{e38}) has the form of two coupled simultaneous
equations:
\begin{equation}
\left(m_N+\omega_{\vec k}\right)\varphi_{\vec k}+h_{\vec k}=E\varphi_{\vec k},
\label{e39a}
\end{equation}
\begin{equation}
m_{V_0}+\int d\vec{k}\,h_{\vec k}\varphi_{\vec k}=E.
\label{e39b}
\end{equation}
Equations (\ref{e39a}) and (\ref{e39b}) are the analog of (\ref{e15}). We solve
(\ref{e39a}) for $\varphi_{\vec k}$,
\begin{equation}
\varphi_{\vec k}=-\frac{h_{\vec k}}{\omega_{\vec k}+m_N-E},\nonumber\\
\label{e40a}
\end{equation}
and substitute the result into (\ref{e39b}) to obtain
\begin{equation}
\int d\vec{k}\,\frac{h_{\vec k}^{2}}{\omega_{\vec k}+m_N-E}=m_{V_0}-E.
\label{e40}
\end{equation}

Next, we define $f(E)$ by
\begin{equation}
f(E)=E-m_{V_0}+\int d\vec{k}\,\frac{h_{\vec k}^2}{\omega_{\vec k}+m_N-E}.
\label{e41}
\end{equation}
The function $f(E)$ is the field-theory analog of the function $f(E)$ that is
plotted in Figs.~\ref{f2} - \ref{f5} and the roots of $f(E)$ are the physical
energy levels of the Lee-model Hamiltonian (\ref{e33}). Note that the first
derivative of $f(E)$ is
\begin{equation}
f'(E)=1+\int d\vec{k}\,\frac{h_{\vec k}^2}{\left(\omega_{\vec k}+m_N-E\right)^2
}.
\label{e42}
\end{equation}

\begin{figure}[b!]\vspace{3.4in}
\includegraphics{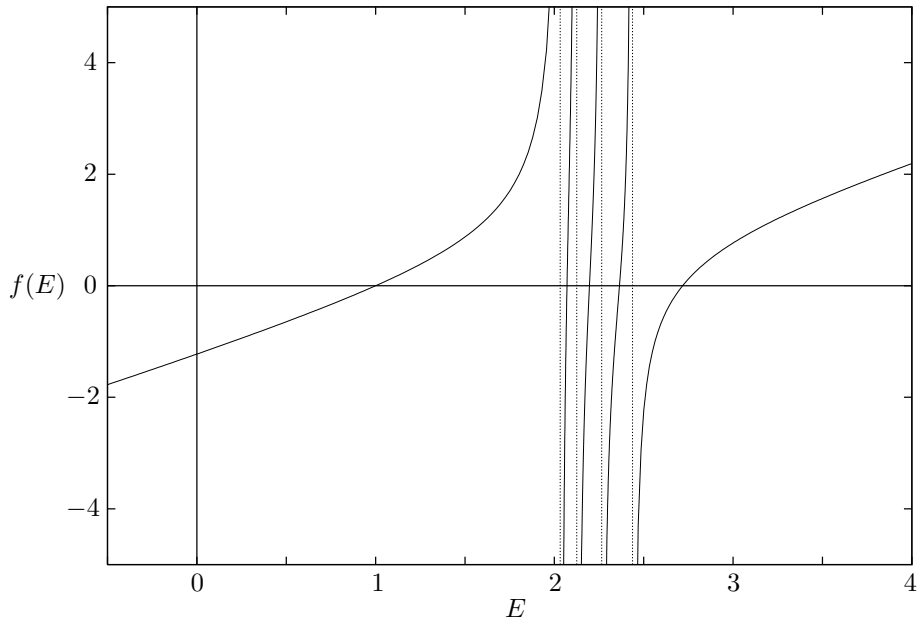}
\caption{Field theoretic version of Fig.~\ref{f2}. The function $f(E)$ in
(\ref{e41}) is plotted as a function of $E$ for the case in which the integral
over $\vec k$ is replaced by a discrete sum over $\vec k$. The energy
eigenvalues of $H$ in (\ref{e33}) are the zeros of $f(E)$. The graph is
constructed for a small real value of $g_0$. The lowest eigenvalue, which
corresponds to the physical $V$ particle, is isolated from the other energies,
which correspond to physical $N$-$\theta$ scattering states. In the continuum
limit in which the discrete sum is replaced by an integral, the physical $N$-$
\theta$ scattering states become dense and form a cut on the real axis in the
complex-$E$ plane.}
\label{f6}
\end{figure}

To show how to use the function $f(E)$ to find the energy levels, we discretize
the integral over $\vec k$ to convert it to a summation. We take $\rho
\left(\omega_{\vec{k}}\right)$ in (\ref{e36}) to be a sharp cut-off function and
we plot in Fig.~\ref{f6} the function $f(E)$ versus $E$ for small real $g_0$.
Observe that there is an isolated energy level that corresponds to the mass of
the physical $V$ particle. We will call this energy level $m_V$. Above this
energy level there is a gap and then a bunch of closely associated energy levels
that correspond to the physical $N$-$\theta$ scattering states. In the continuum
limit the discrete sum over $\vec k$ becomes an integral over $\vec k$, and this
bunch of discrete states becomes a continuous cut on the real axis in the
complex-$E$ plane.

\begin{figure}[b!]\vspace{3.4in}
\includegraphics{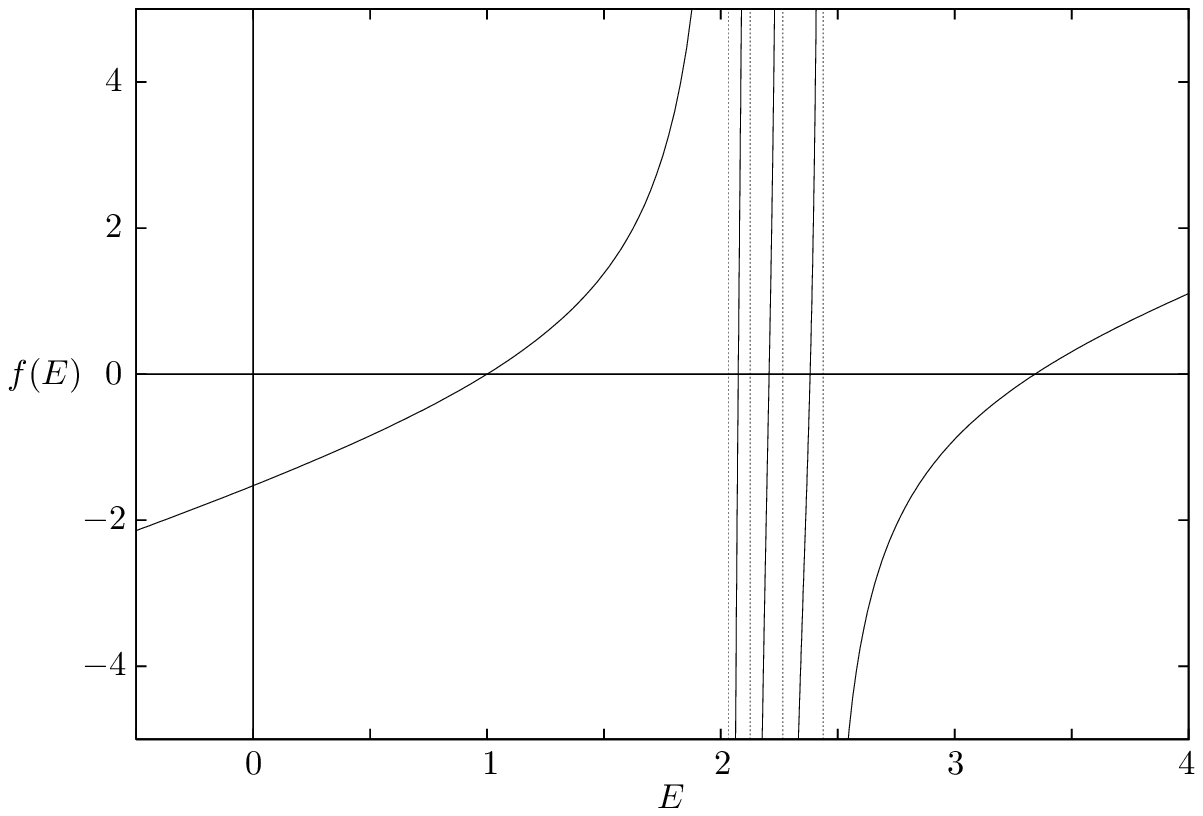}
\caption{Same as in Fig.~\ref{f6} but with $g_0$ having a larger real value.
Observe that as $g_0$ increases, the highest energy physical $N$-$\theta$
scattering state separates from all the rest and moves off towards positive
infinity.}
\label{f7}
\end{figure}

\begin{figure}[b!]\vspace{3.4in}
\includegraphics{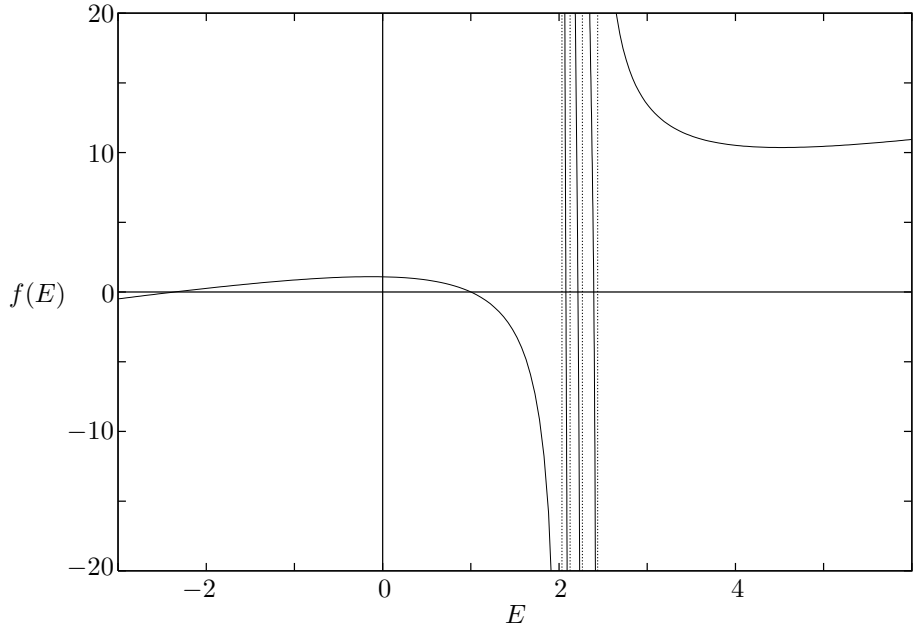}
\caption{Same as in Fig.~\ref{f7} except that the renormalized coupling constant
is larger than its critical value and the unrenormalized coupling constant $g_0$
is imaginary. Observe that the largest zero of $f(E)$ has abruptly become large
and negative. This state is referred to as the ghost state. The ghost state
appears when the Hamiltonian becomes non-Hermitian.}
\label{f8}
\end{figure}

\begin{figure}[b!]\vspace{3.4in}
\includegraphics{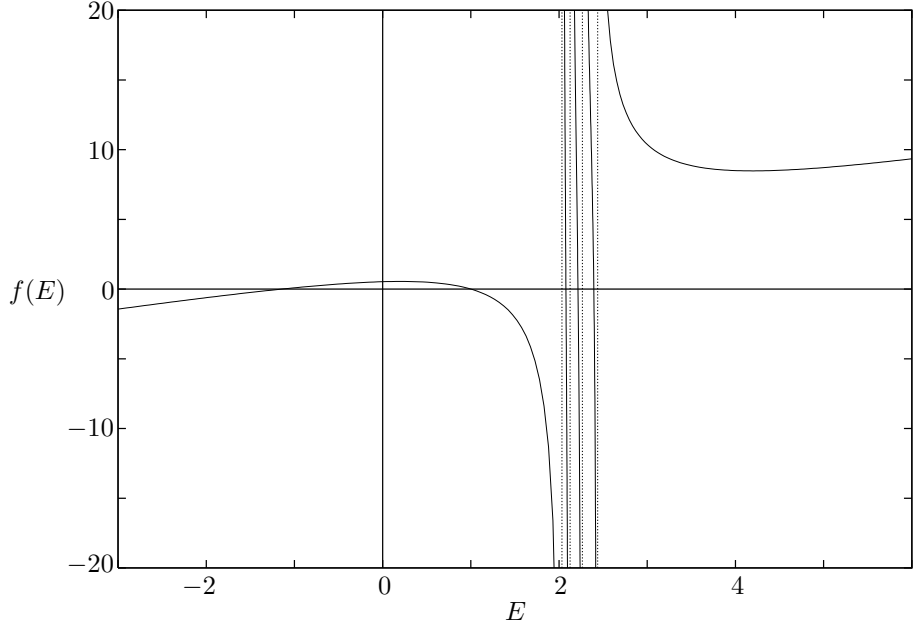}
\caption{Same as in Fig.~\ref{f8} except that the renormalized coupling constant
has a larger value. Observe that the energy of the ghost always lies below the
energy of the physical $V$ particle, but that as $g$ increases, the ghost energy
moves up towards the $V$ energy.}
\label{f9}
\end{figure}

As $g_0$ increases (but still remains real), the highest-energy $N$-$\theta$
scattering state separates from the rest and moves up towards positive
infinity. This separation is illustrated in Fig.~\ref{f7}. Just as the
renormalized coupling constant $g$ increases passes its critical value and the
unrenormalized coupling constant becomes imaginary, the isolated high-energy
state abruptly becomes large and {\it negative} and lies below the physical $V$
state. This jump, which occurs just as the Hamiltonian becomes non-Hermitian, is
characterized by the appearance of the ghost state (see Fig.~\ref{f8}). We will
call this energy level $E_G$. As the renormalized coupling constant continues to
increase, the ghost energy $E_G$ increases towards the $V$-particle energy $m_V$
(see Fig.~\ref{f9}).

Observe that in Figs.~\ref{f6} - \ref{f9}, $E_G$ and $m_V$ lie in the interval
$(-\infty,m_N+m_\theta)$. Also, note that $f'(E_G)>0$ and that $Z_V=f'(m_V)<0$.
When $E>m_N+m_\theta$, $f'(E)<0$. Note also that for any two eigenvalues $E_1
\neq E_2$, we have
\begin{eqnarray}
&&\int d\vec{k}\,\frac{h_{\vec k}^2}{\left(\omega_{\vec k}+m_N-E_1\right)\left(
\omega_{\vec k}+m_N-E_2\right)}\nonumber\\
&&\qquad=\frac{1}{E_1-E_2}\int d\vec{k}\,h_{\vec k}^2
\left[\frac{1}{\omega_{\vec k}+m_N-E_1}-\frac{1}{\omega_{\vec k}+m_N-E_2}\right]
\nonumber\\
&&\qquad=\frac{1}{E_1-E_2}\left[\left(m_{V_0}-E_1\right)-\left(m_{V_0}-E_2
\right)\right]=-1.
\label{e44}
\end{eqnarray}

In the ghost regime, where the Hamiltonian is non-Hermitian, we follow the
approach in Sec.~\ref{s2} and choose the phases $\varepsilon$ of the states in
(\ref{e37}) so that the eigenvalues of the $\cP\cT$ operator are all $+1$. That
is, we require that
\begin{equation}
\cP\cT|E,\vec{p}\rangle=-\varepsilon^*\left(V_{\vec{p}}^\dag-\int d\vec{k}\,
\frac{h_{\vec k}}{\omega_{\vec k}+m_N-E}N_{\vec{p}-\vec{k}}^\dag a_{\vec{k}}^
\dag\right)|0\rangle=|E,\vec{p}\rangle.
\label{e45}
\end{equation}
This requirement implies that $\varepsilon$ must be imaginary:
$\varepsilon^*=-\varepsilon$.

Having chosen the phases of states in this way, we can now calculate the $\cP
\cT$ norms of the states in the $V/N\theta$ sector. To do so we use the general
formula for the $\cP\cT$ norm of the state $|E_1,\vec{p}_1\rangle$:
\begin{eqnarray}
|E_1,\vec{p}_1\rangle^{\cP\cT}\cdot|E_2,\vec{p}_2\rangle&=&\varepsilon_{E_1}
\varepsilon_{E_2}\left(1+\int d\vec{k}\,\frac{h_{\vec k}^2}{\left(\omega_{\vec
k}+m_N-E_1\right)\left(\omega_{\vec k}+m_N-E_2\right)}\right)\delta^{(3)}\left(
\vec{p}_1-\vec{p}_2\right)\nonumber\\
&=&\varepsilon_{E_1}^2f'\left(E_1\right)\delta_{E_1,E_2}\delta^{(3)}\left(
\vec{p}_1-\vec{p}_2\right).
\label{e46}
\end{eqnarray}

For the physical $V$ state and the physical $N$-$\theta$ scattering states, we
have $f'(E)<0$, so by choosing $\varepsilon_E$, we can normalize these states as
\begin{equation}
|E_1,\vec{p}_1\rangle^{\cP\cT}\cdot|E_2,\vec{p}_2\rangle=
\delta_{E_1,E_2}\delta^{(3)}\left(\vec{p}_1-\vec{p}_2\right).
\label{e47}
\end{equation}
However, for the ghost state $f'(E_G)>0$, and we must normalize the state as
\begin{equation}
|E_G,\vec{p}_1\rangle^{\cP\cT}\cdot|E_G,\vec{p}_2\rangle=-\delta^{(3)}
\left(\vec{p}_1-\vec{p}_2\right)
\label{e48}
\end{equation}
Thus the ghost state has a negative $\cP\cT$ norm, while the $V$ state and the
$N$-$\theta$ scattering states have positive $\cP\cT$ norms.

We must introduce the $\cC$ operator in order to repair the negative sign of
the norm of the ghost state. Following the approach in Sec.~\ref{s2} we seek the
$\cC$ operator in the form
\begin{equation}
\cC=e^Q\cP_I,
\label{e49}
\end{equation}
where in this equation $\cP_I$ represents the {\it intrinsic} parity operator
rather than the conventional parity operator $\cP$. The intrinsic parity
operator changes the sign of a pseudoscalar field, but unlike the ordinary
parity operator $\cP$, $\cP_I$ does not change the sign of the spatial variable
in the argument of the field. The exact formula for $\cP_I$ is
\begin{equation}
\cP_I=e^{i\pi(n_V+n_N+n_\theta)}.
\label{e49a}
\end{equation}
Using $\cP$ instead of $\cP_I$ in this formula would give the wrong $\cC$
operator, as we will see in (\ref{e71}) and (\ref{e73}).

In the $V/N\theta$ sector, the conditions $\cC^2=1$ in (\ref{e3a}) and $[\cC,\cP
\cT]=0$ in (\ref{e3b}) require $Q$ to have the form
\begin{equation}
Q=\int d\vec{p}\,d\vec{k}\,\left(\gamma_{\vec k}V_{\vec{p}}^\dag N_{\vec{p}-
\vec{k}}a_{\vec{k}}+\gamma_{\vec k}^*a_{\vec{k}}^\dag N_{\vec{p}-
\vec{k}}^\dag V_{\vec{p}}\right),
\label{e50}
\end{equation}
where the function $\gamma_{\vec k}$ is as yet unknown. In this sector, $Q^2$
can be expressed as
\begin{equation}
Q^{2}=\int d\vec{k}\,|\gamma_{\vec k}|^2\int d\vec{p}\,V_{\vec{p}}^\dag V_{\vec{
p}}+\int d\vec{p}\,d\vec{k}_1\,d\vec{k}_2\,\gamma_{\vec{k}_1}^*\gamma_{\vec{k}_2
}a_{\vec{k}_1}^\dag N_{\vec{p}-\vec{k}_1}^\dag N_{\vec{p}-\vec{k}_2}a_{\vec{k}_2
},
\label{e51}
\end{equation}
where we have ignored the terms which will vanish when applied to a state in
this sector. In this sector, $Q^3$ is proportional to $Q$:
\begin{equation}
Q^3=\int d\vec{l}\,|\gamma_{\vec l}|^2\int d\vec{p}\,d\vec{k}\,\left(\gamma_{
\vec k}V_{\vec{p}}^\dag N_{\vec{p}-\vec{k}}a_{\vec{k}}+\gamma_{\vec k}^*
a_{\vec{k}}^\dag N_{\vec{p}-\vec{k}}^\dag V_{\vec{p}}\right)=\beta^{2}Q,
\label{e52}
\end{equation}
where
\begin{equation}
\beta^2\equiv\int d\vec{k}\,|\gamma_{\vec k}|^2>0.
\label{e53}
\end{equation}
Using these formulas, we can expand $e^Q$ in terms of $Q$ and $Q^2$:
\begin{equation}
e^Q=1+\frac{\sinh\beta}{\beta}Q+\frac{\cosh\beta-1}{\beta^2}Q^2.
\label{e54}
\end{equation}

Next, we impose the condition in (\ref{e3c}) that $\cC$ commutes with $H$. This
gives the formula
\begin{equation}
\left[e^Q,H_0\right]=\left[e^Q,H_I\right]_+.
\label{e55}
\end{equation}
Substituting (\ref{e54}) into (\ref{e55}), we obtain the equations
\begin{eqnarray}
\frac{\sinh\beta}{\beta}\mu_{\vec k}\gamma_{\vec k}&=&2h_{\vec k}+\frac{\cosh
\beta-1}{\beta^2}\left(\beta^2h_{\vec k}+\gamma_{\vec k}\int d\vec{l}
\gamma_{\vec l}^*h_{\vec l}\right),
\label{e56}\\
\frac{\cosh\beta-1}{\beta\sinh\beta}\left(\mu_{\vec{k}_2}-\mu_{\vec{k}_1}\right)
&=&\frac{h_{\vec{k}_2}}{\gamma_{\vec{k}_2}}+\frac{h_{\vec{k}_1}}
{\gamma_{\vec{k}_1}^*},
\label{e57}\\
\int d\vec{k}\,\left(\gamma_{\vec k}+\gamma_{\vec k}^*\right)h_{\vec k}&=&0,
\label{e58}
\end{eqnarray}
where $\mu_{\vec k}\equiv\omega_{\vec k}+m_N-m_{V_0}$. Equation (\ref{e56})
gives
\begin{equation}
\gamma_{\vec k}=\frac{\left(\cosh\beta+1\right)h_{\vec k}}{\frac{\sinh\beta}
{\beta}\mu_{\vec k}-\frac{\cosh\beta-1}{\beta^2}\beta_2},
\label{e59}
\end{equation}
where
\begin{equation}
\beta_2=\int d\vec{k}\,\gamma_{\vec k}^*h_{\vec k}.
\label{e60}
\end{equation}
Substituting (\ref{e59}) into (\ref{e57}), we get
\begin{equation}
\beta_2^*=\beta_2.
\label{e61}
\end{equation}
Thus, $\beta_2$ is real. Combining this result with (\ref{e59}), we find that 
$\gamma_{\vec k}$ is imaginary. Hence, (\ref{e58}) is satisfied.

The numbers $\beta$ and $\beta_2$ satisfy two constraints:
\begin{equation}
\beta^2=-\int d\vec{k}\,\frac{\left(\cosh\beta+1\right)^2h_{\vec k}^2}{\left(
\frac{\sinh\beta}{\beta}\mu_{\vec k}-\frac{\cosh\beta-1}{\beta^2}\beta_2\right)
^2},
\label{e62}
\end{equation}
\begin{equation}
\beta_2=-\int d\vec{k}\,\frac{\left(\cosh\beta+1\right)h_{\vec k}^2}{\frac{
\sinh\beta}{\beta}\mu_{\vec k}-\frac{\cosh\beta-1}{\beta^2}\beta_2}>0.
\label{e62a}
\end{equation}
Let us define
\begin{equation}
E_0\equiv m_{V_0}+\frac{\cosh\beta-1}{\beta\sinh\beta}\beta_2.
\label{e63}
\end{equation}
Then (\ref{e62}) and (\ref{e62a}) become
\begin{eqnarray}
\left(\frac{\sinh\beta}{\cosh\beta+1}\right)^2&=&-\int d\vec{k}\,\frac{h_{
\vec k}^2}{\left(\omega_{\vec k}+m_N-E_0\right)^2},
\label{e64}\\
E_0-m_{V_0}&=&-\int d\vec{k}\,\frac{h_{\vec k}^2}{\omega_{\vec k}+m_N-E_0}.
\label{e65}
\end{eqnarray}
Equation (\ref{e65}) shows that $E_0$ is an {\it eigenvalue} of $H$. From
(\ref{e64}) we have
\begin{equation}
f'(E_0)=1+\int d\vec{k}\,\frac{h_{\vec k}^2}{\left(\omega_{\vec k}+m_N-
E_0\right)^2}=\frac{2}{\cosh\beta+1}>0.
\label{e66}
\end{equation}
We know that only the ghost state has $f'(E)>0$, so
\begin{equation}
E_0=E_G.
\label{e67}
\end{equation}

We can now apply the operator $\cC$ to the eigenstates of $H$:
\begin{eqnarray}
\cC|E,\vec{p}\rangle&=&\varepsilon_E\left(-1+\frac{\sinh\beta}{\beta}\beta_3-
\frac{\cosh\beta-1}{\beta^2}\beta^2\right)V_{\vec p}^\dag|0\rangle\nonumber\\
&&\quad-\varepsilon_E\int d\vec{k}\,\left(\frac{h_{\vec k}}{\omega_{\vec
k}+m_N-E}-\frac{\sinh\beta}{\beta}\gamma_{\vec k}+\frac{\cosh\beta-1}{\beta^2}
\gamma_{\vec k}\beta_3\right)N_{\vec{p}-\vec{k}}^\dag a_{\vec k}^\dag|0\rangle,
\label{e68}
\end{eqnarray}
where
\begin{equation}
\beta_3\equiv\int d\vec{k}\,\frac{\gamma_{\vec k}^*h_{\vec k}}{\omega_{\vec k}+
m_N-E}=-\frac{\beta(\cosh\beta+1)}{\sinh\beta}\int d\vec{k}\,\frac{h_{\vec k}^2}
{\left(\omega_{\vec k}+m_N-E\right)\left(\omega_{\vec k}+m_N-E_0\right)}.
\label{e69}
\end{equation}
If $E\neq E_0$, then from (\ref{e44}) we have
\begin{equation}
\beta_3=\frac{\beta(\cosh\beta+1)}{\sinh\beta}.
\label{e70}
\end{equation}
Substituting this result into (\ref{e68}), we obtain
\begin{equation}
\cC|E,\vec{p}\rangle=|E,\vec{p}\rangle.
\label{e71}
\end{equation}
If $E=E_0$, we have
\begin{equation}
\beta_3=-\frac{\left(\cosh\beta+1\right)\beta}{\sinh\beta}\int d\vec{k}\,
\frac{h_{\vec k}^2}{\left(\omega_{\vec k}+m_N-E_0\right)^2}
=\frac{\beta(\cosh\beta-1)}{\sinh\beta}.
\label{e72}
\end{equation}
In this case (\ref{e68}) is
\begin{equation}
\cC|E,\vec{p}\rangle=-|E,\vec{p}\rangle.
\label{e73}
\end{equation}

Equations (\ref{e71}) and (\ref{e73}) show that it was necessary to use $\cP_I$
and not $\cP$ in the {\it ansatz} for $\cC$ in (\ref{e49}). Had we used the
{\it ansatz} $\cC=e^Q\cP$ we would find that on the right sides of
(\ref{e71}) and (\ref{e73}), the state has the form $|E,-{\vec p}\rangle$.
Thus, the vector $|E,{\vec p}\rangle$ is {\it not} an eigenstate of $\cC$.

Finally, we combine these results with the $\cP\cT$ norm results in (\ref{e47})
and (\ref{e48}) and demonstrate the positivity of the $\cC\cP\cT$ norm:
\begin{equation}
|E_1,\vec{p}_1\rangle^{\cC\cP\cT}\cdot|E_2,\vec{p}_2\rangle=\delta_{E_1,
E_2}\delta^{(3)}\left(\vec{p}_1-\vec{p}_2\right).
\label{e74}
\end{equation}
This shows that in the $V/N\theta$ sector the $\cC\cP\cT$ norms of all states
are positive even when the Hamiltonian is non-Hermitian. Thus, the Lee model
remains unitary in the $V/N\theta$ sector and the ghost state does not interfere
with the unitarity because the Lee-model Hamiltonian becomes $\cP\cT$ symmetric
when it ceases to be Hermitian. Indeed, the use of the term ``ghost'' is
not appropriate because, as we have shown, this state has a positive norm
\cite{last}.

\begin{acknowledgments}
This work was supported by the U.S.~Department of Energy.
\end{acknowledgments}

\end{document}